\documentclass[12pt]{article}
\newcommand{\beq}{\begin{equation}}
\newcommand{\eeq}{\end{equation}}
\newcommand{\bra}{\begin{array}}
\newcommand{\era}{\end{array}}

\newcommand{\al}{\alpha}

\usepackage{latexsym}
\begin{document}
\centerline{\Large\bf Brane Intersections in the Presence of a}
\vspace{1cm}
\centerline{\Large\bf Worldvolume Electric Field}

\vspace{2cm}
\centerline{Rajsekhar Bhattacharyya\footnote{E-address: rbhattac@sun.ac.za} and Jamila Douari\footnote{E-address:douari@sun.ac.za }}
\vspace{1cm}
\centerline{Stellenbosch Institute for Advanced Study}

\centerline{Stellenbosch, South Africa }


\vspace{4.5cm}

\begin{abstract}

The study of brane intersections has provided important insights into a possible non-commutative
structure of spacetime geometry. In this paper we focus on the D1$\bot$D3 system. We compare
the D1 and D3 descriptions of the interesection and search for non-static solutions of 
the D3$\bot$D1 funnel equations in the presence of a worldvolume electric field. 
We find that the D1 and D3 descriptions do not agree. We find time dependent solutions 
that are a natural generalization of those found without the electric field.

\vspace{1cm}


\vspace{0.5cm}
{PACS number(s): 11.25.Uv}
\end{abstract}
\newpage
\section{Introduction}

The study of the low energy dynamics of $D$-branes in string theory, using the Dirac-Born-Infeld
action, has produced many fascinating results [1,2,3,4]. In particular, the 
study of brane intersections has provided deep insights into a possible non-commutative 
structure of the geometry of space-time. These intersections exploit the possibility 
that a $Dp$-brane, coupling to the appropriate fluxes, can be interpreted as $Dp'$-branes of 
lower or higher dimensionality. In this context it is worth mentioning that study of
systems where $D$-strings end on $D(2p+1)$ branes for $p=1,2,3$ has given a concrete
realization of non-commutative geometry, in the form of the co-called non-commutative 
fuzzy funnels [5,6,7]. Recent developments in this direction can be found in [8,9,10,11,12]. In particular, for the $D3 \bot D1$ system, the relevant geometry is that of the fuzzy two sphere. 

One of the interesting aspects to emerge from studies of the $D(2p+1)\bot D1$ systems for 
$p=1,2,3$, is that each system has two descriptions, which may be considered to be dual to each 
other. One description is the $D(2p+1)$-brane worldvolume theory which employs an abelian Dirac-Born-Infeld action with one excited scalar. 
The second description uses the $D$-string 
worldvolume theory. In the absence of a worldvolume electric field, these two descriptions 
are in complete agreement.

Recently time dependent solutions of the $Dp$-brane worldvolume theories have been considered 
in [8,11]. In particular, in [8] space and time dependent fuzzy spheres in the
$D(2p+1)\bot D1$ intersections of IIB string theory have been studied in detail. In the case 
of the $D3\bot D1$ system one finds periodic space and time dependent solutions which can 
be described by Jacobi elliptic functions.

In this paper we consider the $D3 \bot D1$ system with an additional
worldvolume electric field switched on. Our main interest is in the fate of the duality 
mentioned above, and in the fate of the time dependent solutions. We find that duality 
is completely lost; the space time dependent solutions have a nice generalisation. 

\section{Duality in $D3\bot D1$-branes system in presence of electric field}
In this section we study the intersection of a $(p,q)$ string with a D3 brane. We describe this system
as a D3$\bot$D1-brane system, with a worldvolume electric field switched on. For the D1$\bot$D3
system it is well known that the description using the D3 worldvolume theory is in complete
agreement with the description employing the D1 worldvolume theory. Here, by ``complete agreement"
we mean that the two descriptions agree on both the profile of the funnel and the energy of the 
funnel. In this section we are interested in establishing whether a similar result holds even in
the prescence of a worldvolume electric field. We will find that this is not the case. Further, by
carefully considering where the two descriptions are valid we will be able to understand the disagreement.

Our starting point is the Abelian Dirac-Born-Infeld action which describes 
the low energy dynamics of a single $D3$-brane
\beq
S_{DBI}=\int L=-T_3 \int d^4\sigma\sqrt{-det(\eta_{ab}+\lambda^2\partial_a \phi^i \partial_b \phi^i +\lambda F_{ab})}.
\eeq
$F_{ab}$ is the field strength of the $U(1)$ gauge field living on the brane worldvolume and 
$\lambda =2\pi\alpha'=2\pi\ell_s^2 $ with $\ell_s$ the string length. We choose for simplicity to work in static
gauge. 

In the presence of a worldvolume electric field, the energy of the system is
\beq\bra{lll}
\Xi &= T_3 \int d^3\sigma\Big[ 1 +\lambda^2 \Big( \mid \nabla\phi \mid^2 +B^2 +E^2 \Big)\\\\
&+\lambda^4 \Big( (B.\nabla\phi )^2 +(E.B)^2 +\mid E \wedge\nabla\phi\mid^2 \Big) \Big]^{\frac{1}{2}}.\era\eeq
We have traded the field strength for the electric field $F_{0i}=E_i$ and the magnetic field 
$B_r=\frac{1}{2}\epsilon_{rst}F^{st}$. It is straight forward to show that this energy can be brought into the 
following form

\beq\bra{ll}
\Xi &= T_3 \int d^3\sigma\Big[ \lambda^2 \mid \nabla\phi +\stackrel{\rightarrow}{B} +\stackrel{\rightarrow}{E}\mid^2 +(1-\lambda^2 \nabla\phi.\stackrel{\rightarrow}{B})^2-2\lambda^2 \stackrel{\rightarrow}{E}.(\stackrel{\rightarrow}{B} +\nabla\phi)\\\\
&+ \lambda^4 \Big( (\stackrel{\rightarrow}{E}.\stackrel{\rightarrow}{B})^2 +\mid \stackrel{\rightarrow}{E} \Lambda\nabla\phi\mid^2\Big) \Big]^{1/2}.\era\eeq

\noindent
Requiring $\nabla\phi +\stackrel{\rightarrow}{B} +\stackrel{\rightarrow}{E}=0$, $\Xi$ reduces to $\Xi_0\geq0$ 
where
\beq\bra{ll} \Xi_0 &=T_3 \int d^3\sigma\Big[(1-\lambda^2 (\nabla\phi).\stackrel{\rightarrow}{B})^2+2\lambda^2 \stackrel{\rightarrow}{E}.\stackrel{\rightarrow}{E} )\\\\&+ \lambda^4 ( (\stackrel{\rightarrow}{E}.\stackrel{\rightarrow}{B})^2 +{\mid\stackrel{\rightarrow}{E}}\Lambda \nabla\phi\mid^2) \Big]^{1/2}.\era\eeq
For a typical solution of the system 
\beq\phi=\frac{N_m + N_e}{2r}~~~;~~~ \stackrel{\rightarrow}{B}
=\frac{N_m \hat{r}}{2r^2}~~~;~~~\stackrel{\rightarrow}{E}=\frac{N_e \hat{r}}{2r^2}\eeq
we can evaluate $\Xi_0$ to obtain
$$\Xi_0 =T_3 \int d^3\sigma\Big[(1+\lambda^2 \frac{N_m(N_m + N_e)}{4r^4})^2+2\lambda^2 \frac{N^2_e}{4r^4} 
+ \lambda^4 \frac{N^2_m N^2_e}{16r^8}\Big]^{1/2}$$
which can be further reduced to the expression
\beq \Xi_0 =T_3 \int d^3\sigma\Big[1+ \lambda^4\frac{N^2_m[(N_m + N_e)^2+ N^2_e]}{16r^8}+ 2\lambda^2 \frac{N_m(N_m + N_e)}{4r^4}+2\lambda^2 \frac{N^2_e}{4r^4} 
\Big]^{1/2}.\eeq
This provides the description of the system using the D3 brane worldvolume theory.

We now develop a description which employs the D1 brane worldvolume theory. By turning on a worldvolume electric
field, we can use the D-string theory to provide a description of the $(p,q)$ string we are considering. For 
$N_m$ $D$-strings ending on a $D3$-brane with a background $U(1)$ electric field the action becomes [5]
\beq S=-T_1\int d^2\sigma STr \Big[ -det(\eta_{ab}+\lambda^2 \partial_a \phi^i Q_{ij}^{-1}\partial_b \phi^j +\lambda F_{ab})det Q^{ij}\Big]^{1\over2}.\eeq
For the case that we consider, the field strength $F_{ab}= EI_{ab}$ with $I_{ab}$ the $N_m\times N_m$ identity matrix.
Notice that the D-string description involves a non-Abelian theory which is to be contrasted with the Abelian D3-brane
description. Our ansatz for the Higgs fields is
\beq\phi_i =\hat{R}(\sigma)\al_i .\eeq
where $\alpha_i$'s are generators of an $N\times N$ representation of the 
$SU(2)$ algebra$$[\alpha_i,\alpha_j]=2i\epsilon_{ijk}\alpha_k$$
Evaluating the action for this ansatz, we obtain
\beq S=-T_1\int d^2\sigma STr \Big[ (1-\lambda^2 E^2 
+ \al_i \al_i \hat{R}'^2)(1+4\lambda^2 \al_j \al_j \hat{R}^4 )\Big]^{1\over2}.\eeq
Minimizing this action with respect to $R$ for fixed field strength $E$, we obtain the following solution
\beq \phi_i =\frac{\al_i}{2\sigma\sqrt{1-\lambda^2 E^2 	}}\eeq
A comment is in order: one should properly make the ansatz at the level of the equations of motion and not
at the level of the action. We have checked that this solution is indeed consistent with the equations
of motion.

This solution is similar to the dual description of the D3$\bot$D1-system: the scalar field $\phi^i$ describes a fuzzy 
two sphere of radius
\beq R(\sigma)=\lambda\sqrt{\frac{Tr[\phi^i(\sigma)]^2}{N_m}}
=\frac{N_m \pi \ell^2_s}{\sigma\sqrt{1-\lambda^2 E^2 }}\sqrt{1-1/N^2_m}\eeq 
\beq\approx \frac{N_m \pi \ell^2_s}{\sigma\sqrt{1-\lambda^2 E^2 }}.\eeq 
The D3 brane is located at $\sigma=0$ where the radius of the fuzzy sphere diverges. 
Introduce electric displacement $D$ conjugate to $E$, which is, by definition
\beq D=\frac{1}{N_m}\frac{\delta S}{\delta E}=\frac{\lambda E}{g\sqrt{1-\lambda^2 E^2}}=\frac{N_e}{N_m}.\eeq
It is clear that $\frac{\lambda E}{\sqrt{1-\lambda^2 E^2}}=g\frac{N_e}{N_m}$, so $N_e \rightarrow \infty$ when 
$\lambda E \rightarrow 1$. 
Consequently, for sufficiently large but fixed $N_m$, if $N_e \rightarrow \infty$ for fixed $R(\sigma)$ we find 
$\sigma \rightarrow \infty$. This suggests that we should identify $\sigma \leftrightarrow \lambda \phi$ and 
$R \leftrightarrow r$ and which fixes
$$N_e + N_m = \frac{N_m}{\sqrt{1-\lambda^2 E^2}}$$ for $N_e + N_m=N\rightarrow \infty$. 
Thus for large $N$, $\sqrt{1-\lambda^2 E^2} \rightarrow $ some fraction $\frac{m}{n}$ or equivalently 
$\lambda E \rightarrow \sqrt{1-(\frac{m}{n})^2}$ where $0 < \frac{m}{n} <1$. Further from the definition 
of $D$ we have $$N_m\sqrt{(\frac{n}{m})^2-1}=g{N_e}.$$ Thus $N_e \rightarrow \infty$ whenever $m\rightarrow 0$.

Having matched the profiles of the two descriptions, we now ask if the energies of the two descriptions match.
The energy of the $(N_m, N_e)$ system is given by

\beq
\Xi=T_1\int d\sigma \sqrt{N^2_m + g^2 N^2_e} + T_3(1-1/N^2_m)^{-1/2}\int dR 4\pi R^2.
\eeq
We might expect that this energy matches the energy of D3 brane description in the large $N_m$ limit where
both descriptions may be valid. This is indeed the case for the situation with no electric field switched on.

In the presence of an electric field, in the large $N_m$ limit for fixed $N_e$, the energy from the D-string
description, equation (14), becomes $T_1 N_m\int d\sigma $. The energy computed using the
dual D3 description uses (6). We only need to consider the contribution from the second term of the
integrand which is $T_3 \int d^3\sigma  \lambda^2\frac{N_m\sqrt{(N_m + N_e)^2+N^2_e}}{4r^4}$ and using 
$\phi=\frac{N_m + N_e}{2r}$ this can be recast as $T_1 \frac{N_m\sqrt{(N_m + N_e)^2+N^2_e}}{N_m + N_e}\int d\sigma $. 
Again for fixed $N_e$, but in the large $N_m$ limit, $ \frac{\sqrt{(N_m + N_e)^2+N^2_e}}{N_m + N_e}$ becomes 1, so we have 
agreement from the two sides. But we can also take the large $N_m$ limit keeping $N_e/N_m$ fixed at any arbitrary 
$K>0$, then (14) becomes $T_1 N_m \sqrt{1+gK^2}\int d\sigma $ while for (6) again we can only consider the contribution 
from the second term and we find it as $ T_1\frac{N_m\sqrt{(1+K)^2+K^2}}{1+K}\int d\sigma$. Thus in general we have a
disagreement, i.e the presence of electric field spoils the duality between the D3 and D1-brane descriptions of the D3$\bot$D1 system.

To understand why the two descriptions do not match, we will now consider the regimes of validity of the
two descriptions. Both descriptions will receive higher derivative corrections. Our strategy is to estimate
the range of $R(\sigma )$ for which these corrections can safely be ignored. Indeed, the correspodning
analysis in the absence of an electric field shows that the two descriptions have a large overlap in their
domain of validity. In the prescence of an electric field, we find that this is no longer the case.

The higher derivative corrections to the Born-Infeld action arise as terms in the $\ell_s$ expansion of the
low energy effective theory. Due the prescence of the electric field, we find that the leading corrections 
are two derivative terms which take the form $\ell^5_s F_{ab}\partial^a\partial^b \phi$. In the absence of the
electric field the leading corrections is the four derivative term $\ell^6_s (\partial^2\phi)^2$.
We now require $$\ell^5_s F_{ab}\partial^2\phi << \ell^4_s (\partial \phi)^2$$ In the D-string description this
inequality becomes $R(\sigma) >> N_m\ell^2_s E$. For the $D3$ brane description, the inequality is
$R(\sigma) >> \frac{N_m + N_e}{\ell_s E}$. The overlap between the two descriptions in the limit $N_m\rightarrow\infty$ 
limit (for fixed electric field) does not grow. This is in contrast to the case with no electric field where the
overlap in the limit becomes large. This strikingly different behaviour can be traced back to the term
$\ell^5_s F_{ab}\partial^a\partial^b \phi$ whose existence is due to the prescence of the background electric
field.

\section{Non-static $D3\bot D1$ funnels in presence of electric field}
\hspace{.3in} Time dependent solutions of the $D3\bot D1$ funnel equations have been studied 
in detail in [8]. In this section we wish to study the time dependent $D3\bot D1$ funnel 
solutions in the presence of a worldvolume elecric field. Consider $N$ $D$-strings ending on 
a $D3$-brane in the presence of an elecric field. This system has the action (see (7))

\beq S=-T_1\int d^2\sigma STr \Big[ -det(\eta_{ab}+\lambda^2 \partial_a \phi^i Q_{ij}^{-1}\partial_b \phi^j +\lambda F_{ab})det Q^{ij}\Big]^{1\over2}\eeq
in which the field $F_{ab}$ is antisymmetric in the $a$, $b$ indices and 
$F_{\tau\sigma} =E I_{N}$ where $I_{N}$ is the $N\times N$- identity matrix. 
Inserting the space-time dependent ansatz \beq\phi_i =\hat{R}(\sigma, \tau)\al_i \eeq
into the above action, where the $\alpha_i$'s are again the generators of an
$N\times N$ representation of the $SU(2)$ algebra$$[\alpha_i,\alpha_j]=2i\epsilon_{ijk}\alpha_k ,$$
we obtain 
\beq
S=-T_1\int d^2\sigma Str\sqrt{1+\lambda^2 c\hat{R}^{'2}-\lambda^2 c\dot{\hat{R}}^2 -\lambda^2 E^2}\sqrt{1+4\lambda^2 c \hat{R}^4}.
\eeq 
We have made use of the result $\sum^3_{i=1}(\alpha^i)^2=c=N^2-1$.
To express the above action in terms of dimensionless variables we consider the following rescalings
$$
r=\sqrt{2\lambda\sqrt{c}}\hat{R}\phantom{~~~~}\tilde\tau=\sqrt{\frac{2}{\lambda\sqrt{c}}}
\tau\phantom{~~~~}\tilde\sigma=\sqrt{\frac{2}{\lambda\sqrt{c}}}\sigma ~~~~~~\lambda E=e.
$$
The rescaled action is
\beq
\tilde{S}=-\int d^2\sigma \sqrt{1+r'^2-\dot{r}^2 - e^2}\sqrt{1+r^4}.
\eeq 
The conserved (rescaled) energy is 

\beq
T_{\tau\tau}=\Xi=(1+r'^2 - e^2)\sqrt{\frac{1+r^4}{1+r'^2-\dot{r}^2 - e^2}},
\eeq
and the conserved pressure is
\beq
T_{\sigma\sigma}=(1-\dot{r}^2 -e^2)\sqrt{\frac{1+r^4}{1+r'^2-\dot{r}^2 - e^2}},
\eeq
with dots and primes implying differentiation with respect to the rescaled time and space
respectively and to simplify the notation we rename $\tilde\tau$, $\tilde\sigma$ as 
$\tau$, $\sigma$.

We wish to study the purely time dependent solutions of the equation (19) and hence we drop the derivatives of $r$
with respect to $\sigma$. We take the initial condition $r(0)=r_0$ and $\dot{r}(0)=v$. From the 
rescaled energy expression we get 
\beq
\dot{r}^2=\frac{r^4_0(1-e^2)+v^2-r^4 (1-v^2- e^2)}{1+r^4_0}.
\eeq
In terms of \beq  r^4_1=r^4_0+\frac{v^2(1+r^4_0)}{1-v^2-e^2}\eeq we can integrate to obtain
\beq
\int\limits_{0}^{\tau}d\tau'= \sqrt{\frac{1+r^4_0}{1-v^2- e^2}}\int\limits_{r_0}^{r}\frac{dr'}{\sqrt{r^4_1-r'^4}}.
\eeq
This gives
\beq
i\tau\sqrt{\frac{1-v^2-\lambda^2 e^2}{1+r^4_0}}= \int\limits_{r_0}^{r_1}\frac{dr'}{\sqrt{r'^4-r^4_1}}+\int\limits_{r_1}^{r}\frac{dr'}{\sqrt{r'^4-r^4_1}}.
\eeq
Now, using the notation
\beq T=cn^{-1}(\frac{r_0}{r_1},\frac{1}{\sqrt{2}})\eeq the solution can be written as
\beq
r(\tau)=r_1 cn(\bar{\tau}+T,\frac{1}{\sqrt{2}}).
\eeq where $\bar{\tau}=i\tau\sqrt{2}\sqrt{\frac{1-v^2-\lambda^2 e^2}{1+r^4_1}}$.
The interpretation of $r_1$ is now clear: it is the amplitude of the solution, and it controls the
brane radius at the instant in time for which the collapsing speed is zero. Further, (26) implies that we can always cast the time dependence 
of the brane radius in terms of an amplitude and $T$, a $cn$ function, which can be interpreted as the initial phase of the solution.
Using (22), the energy of this solution can be written as $$\Xi=\sqrt{1+r^4_1}.$$

Consider the cases when $v=0$ and $e=0$ separately. For $v=0$ we find (21) becomes
\beq
\dot{r}^2=\frac{r^4_0(1- e^2)-r^4 (1-e^2)}{1+r^4_0},
\eeq
while $r_0=r_1$ and $T$ vanishes.

If $e=0$ (21) becomes
\beq
\dot{r}^2=\frac{r^4_0 + v^2 -r^4 (1-v^2)}{1+r^4_0}.
\eeq We find that now $r_1$ and $r_0$ differ when $T=cn^{-1}(\frac{r_0}{r_1},\frac{1}{\sqrt{2}})$.
The case where both $e=0$ and $v=0$ has been discussed in the reference [8]. 

Consider the equations (21), (27) and (28). We want to examine the solutions in the limit $r_0\rightarrow\infty$, or equivalently,
$r_1\rightarrow\infty$. In this limit we find $\dot{r}^2=1- e^2$ from (21) and (27); in contrast, from (28) and for the case where 
both $e=0$ and $v=0$ we have $\dot{r}^2=1$. Thus, in the presence of a worldvolume electric field, the brane collapses at a speed 
which is less than the speed of light; this speed has been reduced by a factor of $\sqrt{1- e^2}$, and further, the collapsing speed 
does not depend on the initial speed.

In the same way, it is also interesting to study purely spatial dependent solutions $r (\sigma)$; in this case we drop all $\tau$ 
derivatives. Starting from the expression for the conserved pressure, consider $r(\sigma=0)=r_0$ and $r'(\sigma=0)=u$. From (20) we have
\beq
r'^2=\frac{r^4 (1+u^2- e^2) -r^4_0(1-e^2)+u^2}{1+r^4_0},
\eeq
which can be integrated to obtain
\beq
\int\limits_{0}^{\sigma}d\sigma'= \sqrt{\frac{1+r^4_0}{1-u^2- e^2}}\int\limits_{r_0}^{r}\frac{dr'}{\sqrt{r'^4-r^4_1}},
\eeq
where $$r^4_1 =r_0^4 -\frac{u^2 (1+r_0^4 )}{1+u^2 - e^2}.$$  
This can expressed as
$$\frac{\sigma\sqrt{2}r_1 \sqrt{1+u^2 - e^2}}{\sqrt{1+r_0^4 }}=cn^{-1}(\frac{r_1}{r},\frac{1}{\sqrt{2}})-cn^{-1}(\frac{r_1}{r_0},\frac{1}{\sqrt{2}}),$$
$$=cn^{-1}\Big( \frac{\frac{r^2_1}{rr_0}+\frac{1}{2}\sqrt{(1-(\frac{r_1}{r})^4)(1-(\frac{r_1}{r_0})^4)}}{1-\frac{1}{2}(1-(\frac{r_1}{r})^2)(1-(\frac{r_1}{r_0})^2)},\frac{1}{\sqrt{2}}\Big)\eqno(32)$$
If $u=0$ we find (29) becomes $$r'^2 =\frac{r^4 (1- e^2) -r_0^4 (1-e^2)}{1+r_0^4},\eqno(33)$$ $r_1=r_0$ and (32) becomes

$$\frac{\sigma\sqrt{2}r_0 \sqrt{1 - e^2}}{\sqrt{1+r_0^4 }}=cn^{-1}(\frac{r_0}{r},\frac{1}{\sqrt{2}}).
\eqno(34)$$

Further for $e=0$ (29) is modified to $$r'^2 =\frac{r^4 (1+u^2) -r_0^4 +u^2 }{1+r_0^4}\eqno(35)$$ and 
$$r^4_1 =r_0^4 -\frac{u^2(1+r_0^4)}{1+u^2}.$$ The case of $e=0$ and $u=0$ has been studied in [8]. 

Note that for $r_0=u=0$ or for $u^2 =r_0^4 (1-e^2)$, from (29) we find $$r'^2 =r^4 (1- e^2).\eqno(36)$$ This implies
$$\mp r=\frac{1}{\sigma\sqrt{(1- e^2)}\mp \frac{1}{r_0}}.
$$

Thus for particular initial consitions of $r'$ and $r$, we do not find the infinite periodic brane-anti-brane array, but rather a solution with
a decaying $r$. In the presence of an electric field it is attenuated by the factor of $1/\sqrt{1-e^2}$. The equation $r'^2 =r^4 (1- e^2)$ 
provides a generalization of the BPS equation in presence of an electric field. Note that for $e=0$ we get the usual BPS equation $r'^2 =r^4$. 
The BPS equation can be obtained by studying the condition to have an ubroken supersymmetry, which amounts to the condition for the vanishing 
of the variation of gaugino on the world volume of the intersection [8]. A second way to obtain it, is to ask for minimality of (19) for 
$e=\dot{r}=0$, as
$$T_{\tau\tau}=\Xi=\sqrt{(1+r^4)(1+r'^2)}=\sqrt{(r^2\pm r')^2+(1\mp r^2r')^2}$$
and minimality gives $r'^2= r^4$.

For equations (21) and (29), consider the Wick rotation $\tau\rightarrow i\sigma$. This provides interesting results in the cases where $e$, $v$ 
and $u$ are all zero. Here we find (21)$\rightarrow$(29) for $\tau\rightarrow i\sigma$ and $v\rightarrow iu$. Thus in the presence of the electric 
field, the space time dependent solutions generalize nicely. 

Another interesting feature of these solutions is the $r\leftrightarrow 1/r$ duality, which 
arises as the consequence of the invariance of the complex curve under an $r\leftrightarrow 1/r$
automorphism. More precisely,
in the absence of any worldvolume electric field and with $v=0$, we find (21) reduces to 
the equation of a complex curve $s^2=\frac{r^4_0 -r^4}{1+r^4_0}$ where we have defined
$\dot{r}=s$ and $r$ and $s$ are interpreted as complex variables. For this a curve the 
relevant automorphisms have been studied in [8] and a connection to the $r\leftrightarrow 1/r$
duality has been made. Here we find a similar duality. Towards this end, we study 
an automorphism of the curve (21) for $e\neq 0$ and $v\neq 0$. After defining $s=\dot{r}$ we 
find
$$s^2=\frac{r^4_0 (1-e^2)+v^2-r^4 (1-v^2- e^2)}{1+r^4_0}.
\eqno(37)$$
The automorphism acts as $r\rightarrow R=\frac{1}{r}$, $r_0\rightarrow R_0=\frac{1}{r_0}$, $s\rightarrow \tilde{s}=\frac{is}{r^2}$, $v\rightarrow v'=-ivR_0^2$. This automrphism acts at 
fixed $e$. For $v=0$, $r\rightarrow R=\frac{r^2_0}{r}$, $s\rightarrow \tilde{s}=\frac{isr^2_0}{r^2}$, is still a good automorphism of (37).

\section{Conclusion}
\hspace{.3in}
In presence of an electric field the duality between the D1 and D3 description is no longer valid. We have further argued that the
D-string description breaks down. This is perhaps not surprising. Indeed, in this limit [13] have argued that the effective tension
of the string goes to zero. Thus, excited strings modes will not be very heavy compared to massless string modes and one might
question the validity of the Dirac-Born-Infeld action which retains only the massless modes.

Further in the presence of a world volume electric field, space time dependent solutions can be
generalised nicely and for $r_0=0$ we observe brane collapses with a speed less than that of 
light. We have also obtained a generalisation of the BPS solution. An automorphism
of our solutions relevant for the $r\leftrightarrow 1/r$ duality has been discussed.  

\vspace*{1cm}
{\bf Acknowledgment}

\vspace*{0.5cm}

We would like to thank Robert de Mello Koch for very pleasant and helpful discussion and for the comments on the manuscript.

\end{document}